%% file: spieproc_corr.tex
\documentclass [a4paper]{spie}
\usepackage {graphicx}
\usepackage {color}

\title{ Investigation of Residual Blaze Functions in Slit-Based Echelle Spectrograph}

\author{Petr \v{S}koda\supit{a}, Brankica \v{S}urlan\supit{b} and Sanja Tomi\'c\supit{b}
\skiplinehalf
\supit{a} Astronomical Institute Academy of Sciences, Fri\v{c}ova 298, 251 65 Ond\v{r}ejov, Czech Republic; \\
\supit{b} Department of Astronomy, University of Belgrade, Studentski trg 16, 11000 Belgrade, Serbia 
}

\authorinfo{Please send correspondence to P. \v{S}koda: E-mail skoda@sunstel.asu.cas.cz, telephone +420 323 620 361}
\begin{document}
\maketitle

\begin{abstract}
We have studied the Residual Blaze Functions (RBF) resulting from division of
individual echelle orders by extracted flat-field in spectra obtained by
slit-fed OES spectrograph of 2m telescope of Ond\v{r}ejov observatory, Czech
Republic. We have eliminated the dependence on target and observation
conditions by semiautomatic fitting of global response function, thus getting
the instrument-only dependent part, which may be easily incorporated into data
reduction pipeline. The improvement of reliability of estimation of continuum
on spectra of targets with wide and shallow lines is noticeable and the merging
of all orders into the one long spectrum gives much more reliable results.
\end{abstract}


\section{Introduction}

The requirement of a modern astrophysical analysis, to study the wide range of
the object spectrum in the greatest details, or to observe its variability in
many spectral lines at the same time, lead naturally to the usage of
cross-dispersed echelle spectrographs.

Although the reduction of the echelle raw frames is quite straightforward and a
number of automatic reduction pipelines has been in use for many instruments,
there remains still unsolved the problem of the reliable normalization of
individual extracted spectral orders before merging into one long continuous
spectrum.

Due to its inherent nature the echelle blaze function has to be removed from
all orders with the immense precision, otherwise the strong ripple patterns
appear on the merged spectrum. They   make difficult the fitting of the
continuum an hence influence the reliability of the measurement of equivalent
widths of wide shallow lines, the placement of wings of hydrogen lines of hot
rapidly rotating stars, or the studies of diffuse interstellar bands.

\section{Basic Reduction of Echelle Spectra}

\vspace{1ex}

\subsection{Characteristic Features of Echelle Spectrographs}

Almost all modern high-dispersion spectrographs used today are the
cross-dispersed echelle spectrographs giving the very high angular dispersion
and the extensive wavelength coverage at the same time. The individual spectral
orders (typically 30 to 100), representing short piece of the spectrum (several
tens of \AA), are imaged at the rectangular CCD chip as a slanted parallel
strips, sometimes even slightly curved. This breeds, however, the large complex
of problems concerning the data reduction as the correct echelle data reduction
is very sensitive to  even very small errors or omissions.

\subsection{General Reduction Techniques}

The basic principles of the reduction is described in a number of places
\cite{doecslit,1988PASP..100.1572G,1994PASP..106..315H,2000vnia.book.....H}.
There are also detailed cookbooks for reduction using particular package
(mainly IRAF\cite{licktechrep74, 2002A&A...383..227E}).

The general reduction requires a number of steps to remove the detector
features as in all CCD processing. Here comes the bias removal and overscan
correction and (if necessary) dark current elimination.  Then there might
follow classical flat-fielding using the original 2-dimensional flat image, but
it is generally not done due to problems with blaze function (see later).  The
quite difficult task in this part is the removal of fringing that disturbs the
spectra in the IR region on the thinned chips.  The preferred way is the
flat-fielding on the extracted 1-dimensional spectra, that also removes the
main part of blaze function.

On this cleaned flat field frame the echelle cross-order positions are found by
looking through central cut of the frame, perpendicular to dispersion, and then
the cross-order centers are traced towards both ends. Then a bilinear
polynomials are fitted through all cross-order center positions.  In this step
the background, containing the scattered light, has to be removed.  Sometimes
the instrument bias elimination is postponed to this moment. The scattered
light can be found by taking the median values of region between echelle orders
and fitting a 2-dimensional surface through it, that is then subtracted from
the data. Although it might work well for certain instruments with the large
gaps between the orders, for the instruments with densely packed orders the
physical model of the scattering has to be applied. This is very difficult and
computing demanding task
\cite{1994PASP..106..315H,1986A&A...168..386G,licktechrep74,
1988PASP..100.1572G} but it has great importance even for the very precise
space spectrograph STIS \cite{2000AJ....119.2481H}.

Now it can begin the extraction of the flux from the stellar orders following
the traces of cross-order center positions --- usually using traces from flat
field orders. The extraction is done with the optimal variance-weighted methods
\cite{1986PASP...98..609H,1990PASP..102..183M} eliminating the cosmic rays at
the same time.  There are still attempts to improve this well-proved algorithms
\cite{2002A&A...385.1095P}.

The critical part is now the wavelength calibration by fitting 2-dimensional
polynomials through the measured position of a hundreds of sharp lines of the
comparison arc spectrum.  The availability of more precise instruments also
needs the re-analysis of standard calibration algorithms
\cite{1998A&AS..128..409D}.

Although quite complicated, this part of echelle reduction is quite
straightforward and the automatic pipelines can be used to reduce the data. We
end up with separate echelle orders with associated wavelength scale.

Such  data are  very useful for measurement of radial velocities of many lines
(used in cross-correlation techniques \cite{2000PASP..112..966S}  on late-type
stars with many sharp lines) or for study of abundances using the equivalent
width of many deep sharp lines of heavy elements \cite{1998A&A...338..151K}.

\subsection{The Merging of Individual Echelle Orders}

The disadvantage of echelle spectra is the very short spectral span of
individual orders (typically less than 100~\AA). The work on individual orders
can be done for late-type stars with narrow lines, where the sufficient
pseudo-continuum windows are seen in each order.  For hot and rapidly rotating
stars, where  hydrogen and helium lines are of great interest, the considerably
wide line profile is spread over several echelle orders. Therefore the merging
of several orders  is required to see the whole profile together with the
surrounding pseudo-continuum. Although the merging can be done quite easily
after rebinning of orders to the same wavelength grid (preferably in $\log\,
\lambda$), by weighted averaging  of the overlapping regions (and cutting the
very edges where the high noise is present), this step requires extremely
precise handling of data to get reliable results. The problem is caused by the
behavior of the echelle blaze function.

\subsection{The Blaze Function}

The most serious problem in the reduction, that has not been yet solved
satisfactorily, is the removal of the grating blaze function. The blaze
function changes the intensity of the spectrum inside each order and thus
modulates strongly the shape of the stellar continuum.  In each order, the
intensity of signal steeply rises from one edge to the center of frame and
falls down to the other edge. 

\subsubsection{Methods of blaze function removal}

There are several methods of the blaze function removal.  One of the most
promising is the model of its theoretical shape.  From the diffraction theory
the blaze function should behave like a sinc-squared function of the
spectrograph construction parameters (the angle of incidence, angle of blaze,
grating grooves width and spacing). If knowing these parameters precisely, a
model of blaze may be constructed. It is, however, only an approximation, as
the real blaze is not produced by an ideal grating, but other construction
features have its influence as well.

This method developed by Barker\cite{1984AJ.....89..899B} was applied to the
high-dispersion spectra from IUE satellite and implemented later by Cassattella
in IUE NEWSIPS pipeline. Although it was quite successful as the first
approximation of ripple correction and may be used to remove the largest part
of intensity modulation before using sophisticated differential methods
\cite{1997PASP..109..868V}, it will not completely remove the influence of the
blaze from data.

Other method is the usage of the spectra of the spectrophotometric standard
taken at about the same time as the primary target. The identically processed
data of the standard are then divided by its absolute flux and the ratio is
used to get the flux of the object.  Due to its dependence on the extinction
and the requirement of a sufficiently bright standard in the target  proximity
this method is used mainly in space instruments --- e.g. for reducing HST STIS
data. Even here some problems with order merging remains\cite{stis2002}.

The widely used method at most modern spectrographs is the division by the
extracted flat field  spectrum. This method should give the best results,
mainly for fiber-fed spectrographs, where the cross-order profile of the star
and flat is the same, provided that their blaze function is identical.  In
practice, however, this is not the case, and the blaze function shows small
variations of still unknown origin. Some experiments with the fiber-fed
spectrographs give clues to the suspicion of the influence of flat-field
calibration units and the influence of fiber itself. But some part of the
changes is probably intrinsic to the grating theory itself (e.g. polarization
change).

There are, thus, attempts to remove the blaze function by fitting the smooth
surface through only the extracted stellar  data in pixel-order space, using
only the orders where the continuum is present.  In this case, the flat field
is used only for adjusting the CCD pixels sensitivity --- first it is smoothed
still in original 2-dimensional frame.  This method, however, fails in the blue
region on the spectra of hot or rapidly rotating stars where almost every order
is contaminated by wide Balmer lines.

\subsubsection{Blaze function instability}

The tiny changes in position of the orders as well as in the cross-order
profile shape are commonly observed between flat field and stellar spectrum
even in case of the well stabilized instrument in protected environment like
FEROS \cite{feros2002} or ELODIE \cite{2002A&A...383..227E}.

Corresponding small shift of blaze function then causes the tilt of individual
flat-fielded orders and introduces intensity jumps between the overlapping
edges of successive orders. If such  data is blindly merged uncorrected (as in
the automatic pipeline), the strong ripple-shaped periodic undulations with the
period corresponding to free spectral range occur on the continuum shape.  The
extracted flat-field spectra mutually divided show clearly the time-dependent
variations of order of several percent.  \cite{1998adass...7..312D}.  This
behavior is  confirmed by our experience with the HEROS and FEROS data as well
\cite{2003adass..12..415S}.

The blaze function time-dependent variations were also reported on HST STIS
spectrograph by Bowers and Lindler \cite{stis2002}. They have the suspicion,
that the blaze changes are caused not by the grating movement but by the
changes of the grating surface (size and position of individual grooves).

\subsubsection{The influence of the fiber or slit on blaze function}

Although the light from the flat field and star goes through the same slit and
the same narrow decker, the illumination pattern of the flat is generally
different from the stellar one, it changes with the seeing and depends on the
precision of the guiding. The influence of these centering errors on radial
velocity shifts was studied in detail\cite{1997PASP..109..868V}, but it will
cause the micro-shifts of the order positions as well.  In case of fiber-fed
spectrographs this effect should not be present, as is generally believed.
There are, however, evidences that it is not true at the level of precision we
need.

There has been proved  the influence of the optical fiber on the transported
light which  will slightly change the position of spectral orders on the
detector.  Even the double fiber scrambler (between the two pieces of chopped
fiber are inserted micro-lenses) used for the precise RV measurement (mainly in
search for extrasolar planets) does not help to eliminate all the blaze
function variations. One of the instruments equipped with the fiber scrambler
is ELODIE \cite{1998bdep.conf..324Q}, but spectra from it had to be
artificially corrected before merging orders as well
\cite{2001A&A...369.1048P}. The same problems are present in spectra from more
advanced high precision instrument SOPHIE (Ilovaisky 2007, private
communication).

\subsection{The Problem of Continuum Normalization}

Although most of the reduction procedures give quite satisfactory results in
blaze function removal, there still remain small systematic errors in the
unblazed spectra.  An automatic merging of orders then may result in periodic
ripple disturbances in the shape of the apparent stellar continuum.  A typical
way to remove them is the manual fitting of a sufficiently smooth spline
function through the parts of the merged  spectrum where the continuum seems to
be present.  However, for the wide lines with large wings it is very
difficult to do the normalization by spline interpolation between neighboring
continuum windows if the ripples are present. 

\section{Merging of OES Spectra}

\subsection{Ond\v{r}ejov Echelle Spectrograph (OES)}

\begin{figure}[h]
\begin{center}
\includegraphics[width=13cm,height=8.5cm]{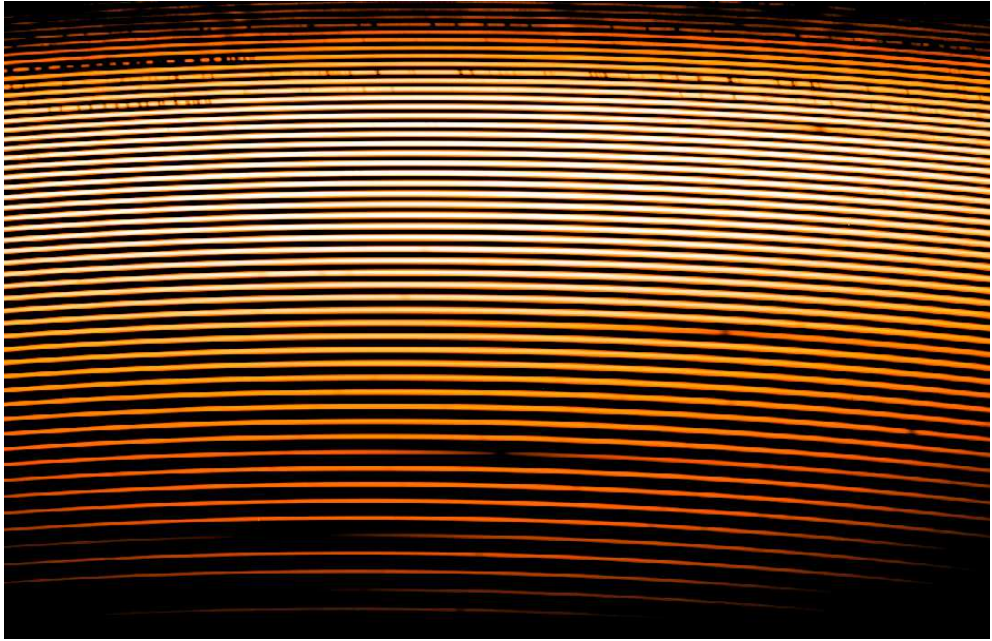}\\
\caption{Part of OES echelle-gram of Vega.}
\label{OESgram}
\end{center}
\end{figure}

The OES\cite{2004PAICz..92...37K} is the slit-fed prism cross-dispersed echelle
spectrograph  developed at the Stellar department of the Astronomical Institute
of the Academy of Sciences of the Czech Republic and installed at the coud\`e
focus of 2m telescope of the Ond\v{r}ejov observatory. In its current setup it
can cover  spectral range from 3750\AA--9500\AA\ in 62 orders. Due to various
construction limitations there is a lack of order overlap at wavelengths longer
than about 6000 \AA.  This together with strongly curved orders, tilted
spectral lines and high level of inter-order scattered light makes reduction
extremely complicated.  Moreover, there is some vignettation seen at the edges
of the orders.  The example of the echelle-gram is given on the
Fig.~\ref{OESgram}.

\subsection{Residual Blaze Function}

\begin{figure}[h]
\begin{center}
\includegraphics[angle=-90,width=0.45\textwidth]{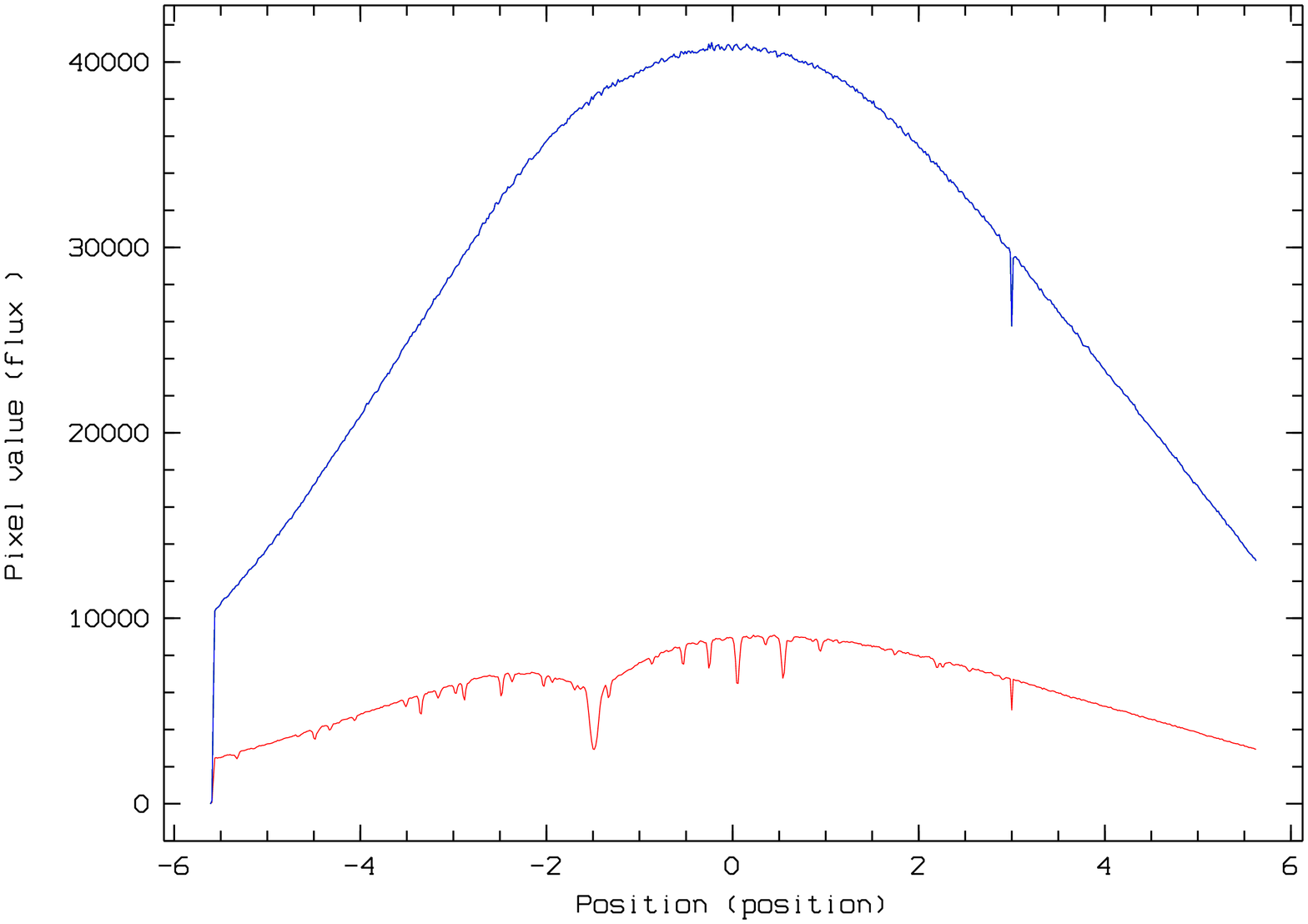}
\includegraphics[angle=-90,width=0.45\textwidth]{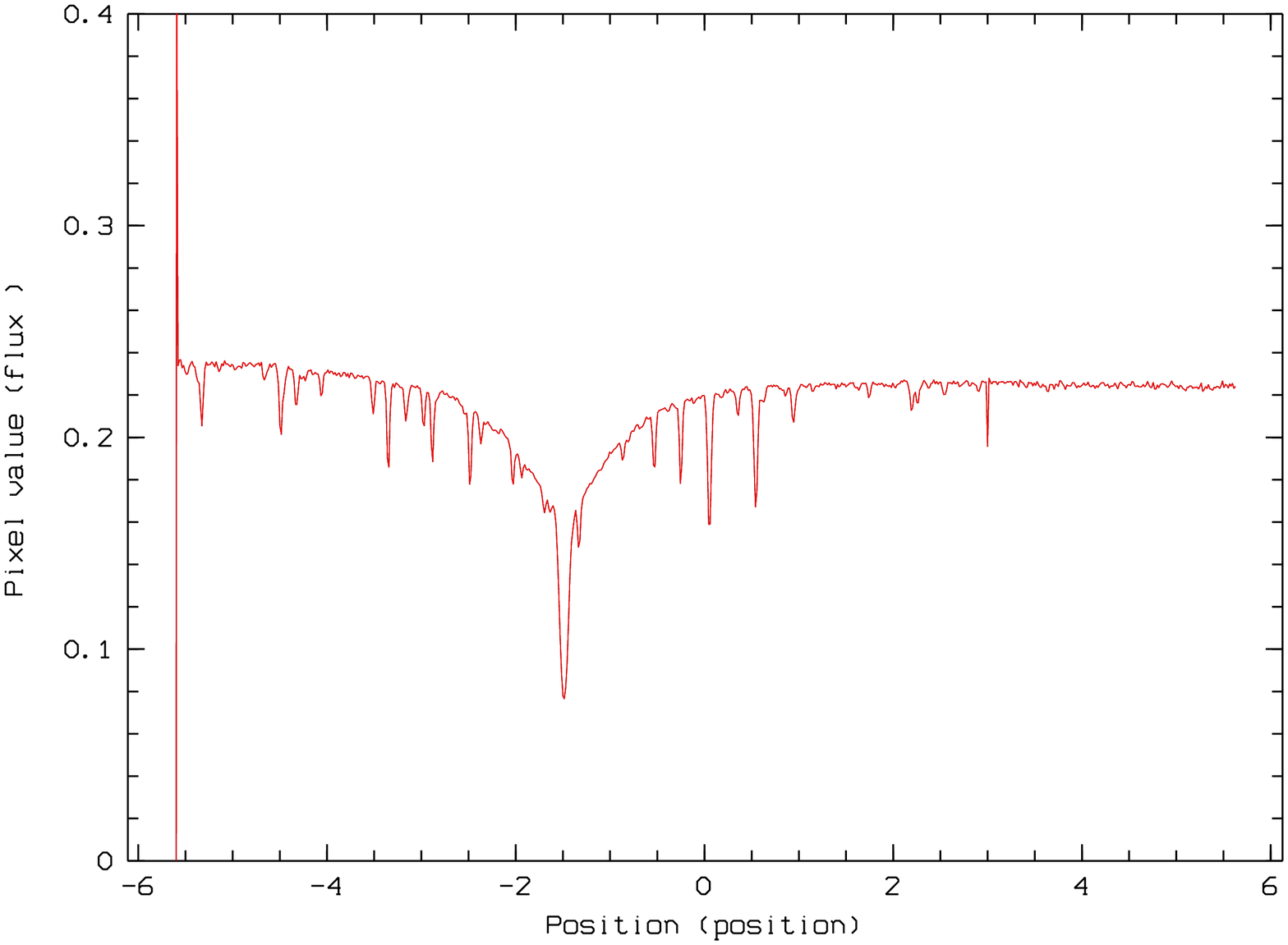}\\
\caption{HEROS: \textbf{Left:} Extracted orders of star and flat.
\textbf{Right:} Stellar order after division by extracted flat.}
\label{HEROScomp}
\end{center}
\end{figure}

In general the  blaze function of stellar exposures on slit-fed echelle
spectrographs is quite different from blaze function of the flat field spectrum
due to different character of slit illumination. So the unblazing by direct
division of the extracted stellar continuum  by the extracted flat field does
not remove the shape of blaze function completely and some residual structure
remains superimposed on the stellar continuum.  For such a curve we are using
the term Residual Blaze Function (RBF).  In the ideal case it should be the
line of constant value corresponding to the ratio of intensity of stellar and
flat field exposures. This is better fulfilled by fiber-fed spectrographs as
the illumination of flat and star is almost the same due to the same size of
fiber entrance with a micro-lens. The example of the behavior of one spectral
order on fiber spectrograph HEROS is given on Fig.~\ref{HEROScomp}.

\begin{figure}[h]
\begin{center}
\includegraphics[width=6.7cm,height=5.7cm]{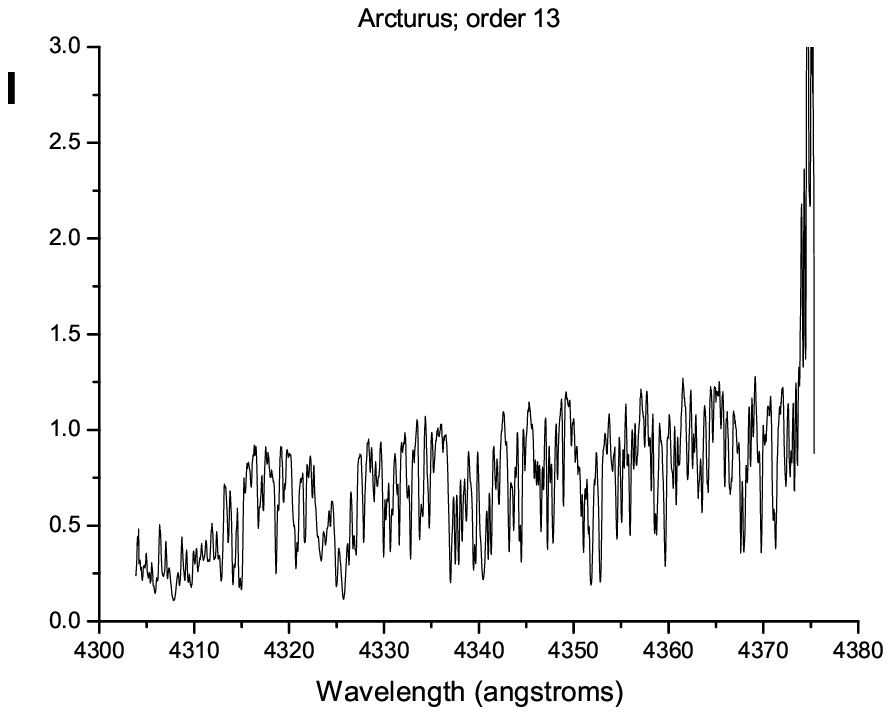}
\includegraphics[width=6.7cm,height=5.7cm]{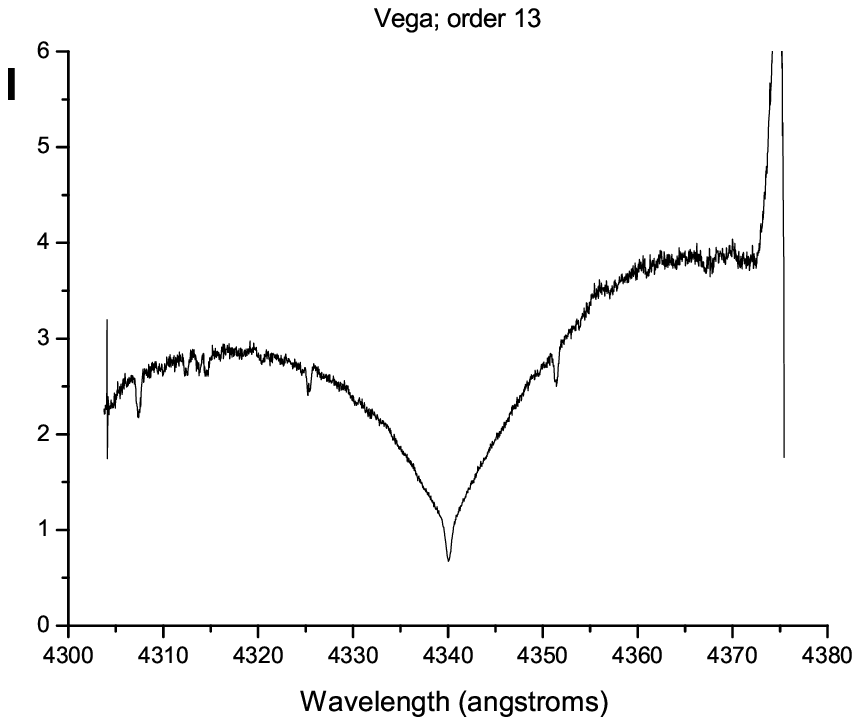}\\
\caption{OES: unblazed order 13 with line $H_{\gamma}$.  \textbf{Left:} Arcturus.
\textbf{Right:} Vega}
\label{BalmerRBF}
\end{center}
\end{figure}

\begin{figure}[h]
\begin{center}
\includegraphics[width=6.7cm,height=5.7cm]{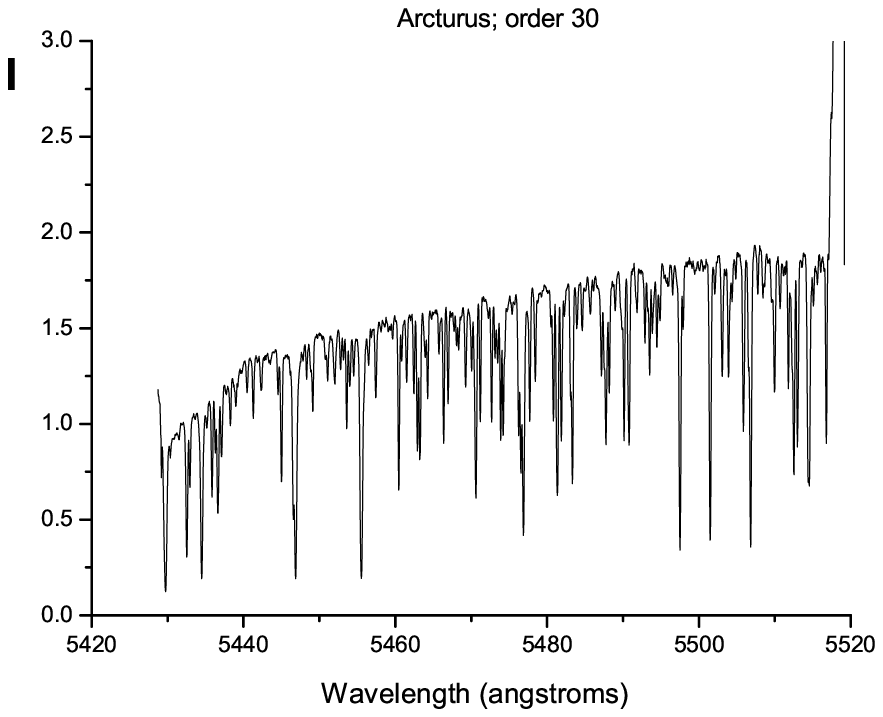}
\includegraphics[width=6.7cm,height=5.7cm]{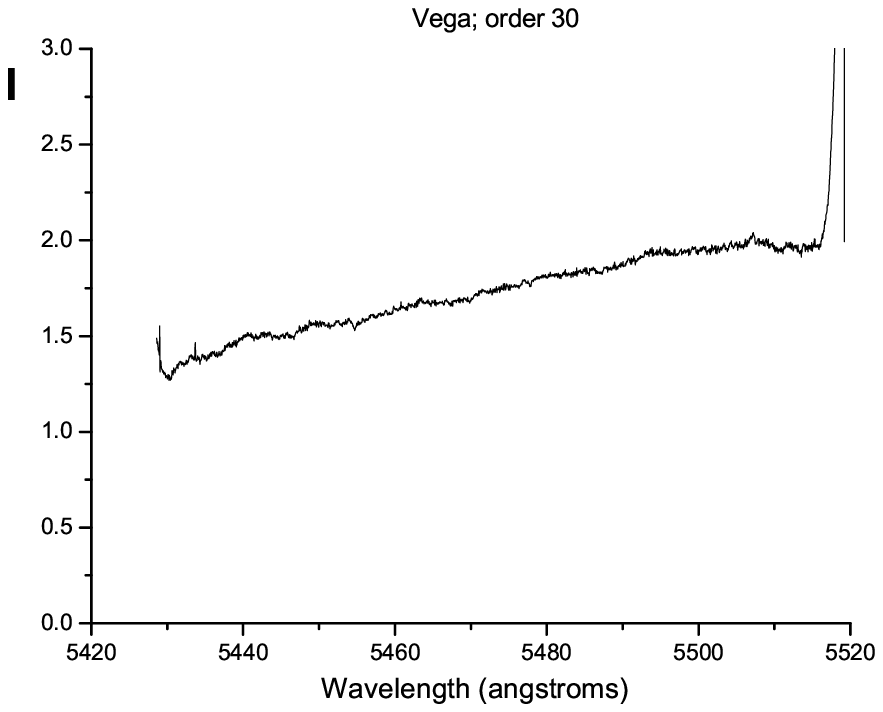}\\
\caption{OES: unblazed order 30 with continuum. \textbf{Left:} Arcturus 
\textbf{Right:} Vega}
\label{contRBF}
\end{center}
\end{figure}

However, the shape of such orders in case of OES is more complicated mainly
influenced by vignettation due to incompatibility of illumination structure.
Examples of   orders containing line ${H_\gamma}$ and ones with obvious
continuum for case of  Vega and Arcturus are given on Fig.~\ref{BalmerRBF} and
Fig.~\ref{contRBF}.

\subsubsection{Estimation of the Residual Blaze Function}

As the true RBF is unknown (we do not see the true continuum due to wide lines
and blends), we have to estimate its shape using comparison with other spectrum
with already known RBF.  The best case is the spectrum of the same target with
approximately same resolution that is already continuum normalized (the
continuum reference spectrum).  If the resolution is not similar, the high
resolution spectrum has to be convolved with sufficiently wide Gaussian profile
to reach comparable lower  resolution. One good option is the spectrum from
single order spectrograph, where the continuum is easily obtained.

\subsubsection{Preparation of continuum normalized reference spectrum}

In the ideal case of several bright stars, there are available their normalized
merged spectra either in web archives (e.g. ELODIE
archive\cite{2001A&A...369.1048P}) or published spectral atlases.  For our case
we used the high resolution atlas of Arcturus by Hinkle et
al.\cite{2000vnia.book.....H} and Vega atlas of Takeda et al.
\cite{2007PASJ...59..245T}

We also tried to rectify the spectra of Vega and Arcturus order by order
ourselves to preserve the resolution.  The rectification polynomials were
applied iteratively in custom program  SPEFO\cite{1996ASPC..101..187S},
checking the consistency of order overlaps and finally comparing visually the
merged rectified spectrum with the synthetic spectrum.  

We used synthetic stellar spectra with temperatures approximately corresponding
to temperature of Arcturus (Teff=4300 K) and Vega (Teff=9300 K).  After that we
used the "{\texttt {sYnt-Rotate}}" command of SPEFO program, where we put
corresponding rotation velocity of Arcturus ($v \sin i$=2 km/s) and Vega ($v
\sin i$=25 km/s). After required rotational broadening we used those synthetic
spectra in further visual comparisons.

\subsubsection{Determination of Residual Blaze Function}

\begin{figure}[h!]
\begin{center}
\includegraphics[width=0.45\textwidth]{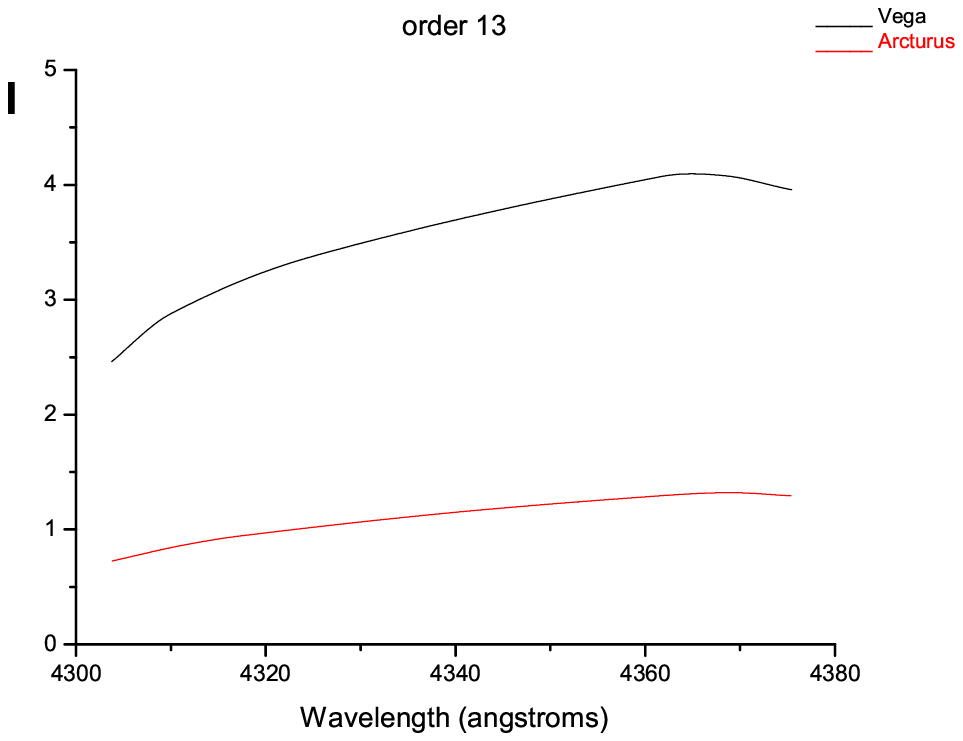}
\includegraphics[width=0.45\textwidth]{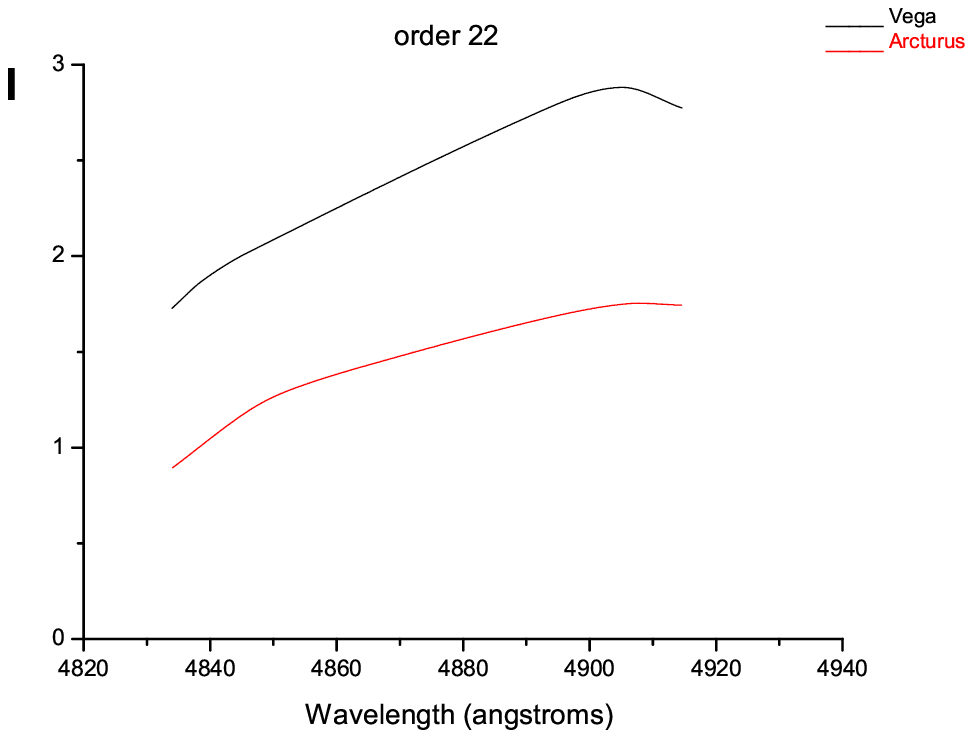}\\
\caption{Comparison of RBFs in two orders of Arcturus and Vega. \textbf{Left:} Order 13.  \textbf{Right:} Order 22. }
\label{RBFord}
\end{center}
\end{figure}

\begin{figure}[h!]
\begin{center}
\includegraphics[width=11.5cm,height=8.5cm]{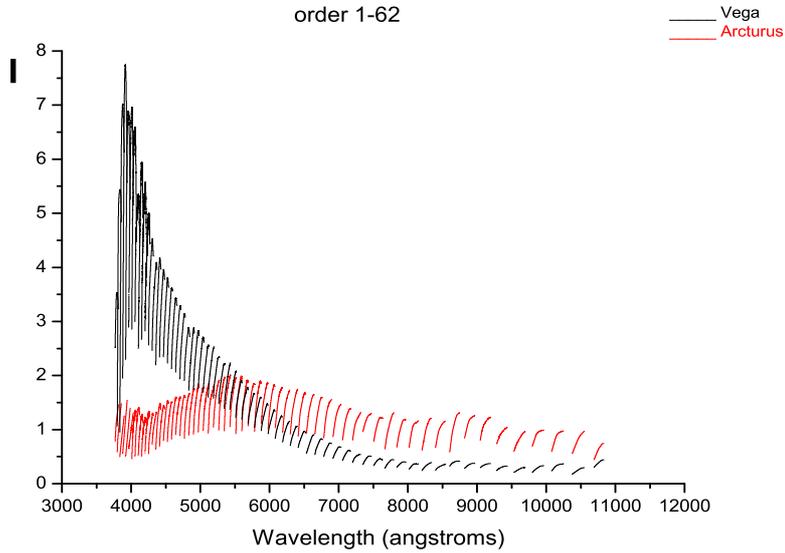}\\
\caption{Comparison of global shape of residual blaze
functions  for Arcturus and Vega.}
\label{RBF}
\end{center}
\end{figure}

After dividing non-rectified spectrum (individual orders) of Arcturus and Vega
by  rectified ones (continuum master reference), we get the   shape of residual
blaze function for each order. Following figures show comparison of shape of
residual blaze functions in two different orders (containing Balmer lines
${H_\gamma}$  and ${H_\beta}$) of Arcturus and Vega (Fig.~\ref{RBFord}).  The
global view  of  RBFs in all orders can be seen on Fig.~\ref{RBF}.

\subsection{The Global Sensitivity Function} 

As is seen on Fig.~\ref{RBF}, the RBFs are following some smooth curve
different for Arcturus and the Vega.  We  suppose that this function, that we
call the Global Sensitivity Function (GSF), depends on the ratio of energy
distributions of the target and the flat field lamp and  on the relative
sensitivity of CCD detector in given spectral region.

Given constant source of flat field lamp and the stable CCD detector, we can
suppose GSF is only dependent on the target color and extinction (spectral
energy distribution- SED).  Color changes are expected to be dependent on the
extinction and hence the zenith distance of the target, as well as on seeing,
atmospheric differential refraction and other observation-dependent parameters.
That is the reason why we have chosen Vega and Arcturus as representants of two
different SEDs (cold and hot star) and  extinction (different zenith distance
at the time of observation).

\subsubsection{Fitting of global sensitivity function}

\begin{figure}[h]
\begin{center}
\includegraphics[width=11.5cm,height=8.5cm]{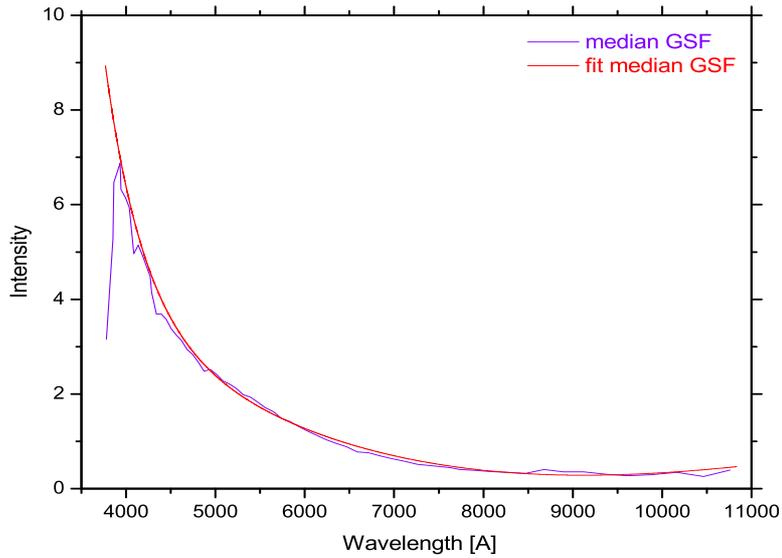}\\
\caption{Fit of GSF through median points in Vega orders}
\label{median_GSF}
\end{center}
\end{figure}

To find some representative scaling point-to-point  on each order one could use
the approximate center of each order. We have used, however, the median of
distribution of values in every order. The median better represents some
``middle'' point for case of strongly curved RBF. The orders contaminated by
wide Balmer lines have even this estimate wrong, but the remaining points (with
possible elimination of known contaminated orders) are easily fitted by smooth
polynomial while rejecting deviant points in an iterative procedure
(Fig.~\ref{median_GSF}).

The knowledge of target GSF  then can prepare rectified individual orders even
before merging, but the normalization can be done easily on merged spectrum as
well. 

\subsection{The Intrinsic Residual Blaze Function}

\begin{figure}[h]
\begin{center}
\includegraphics[angle=90,width=0.45\textwidth]{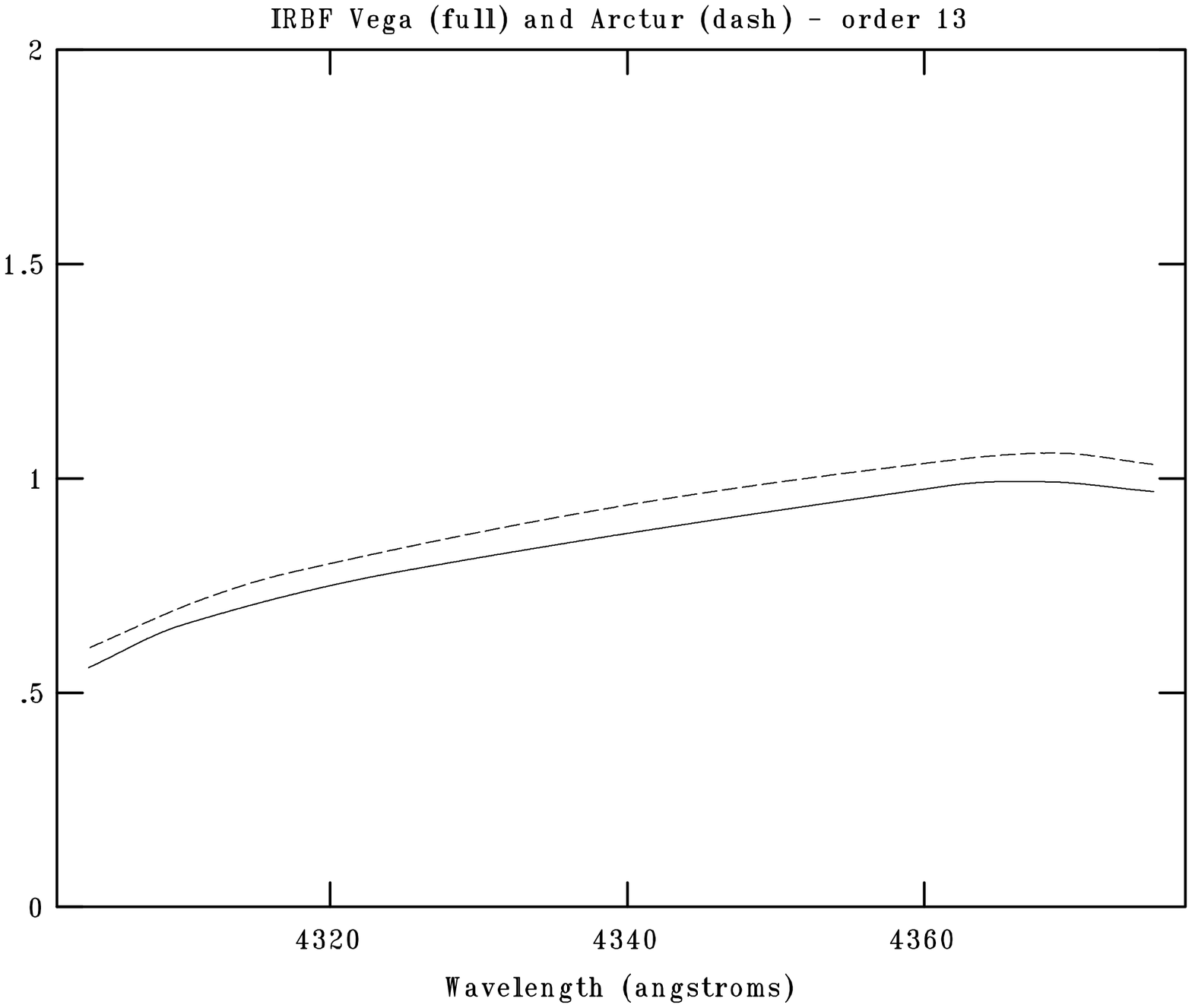}
\includegraphics[angle=90,width=0.45\textwidth]{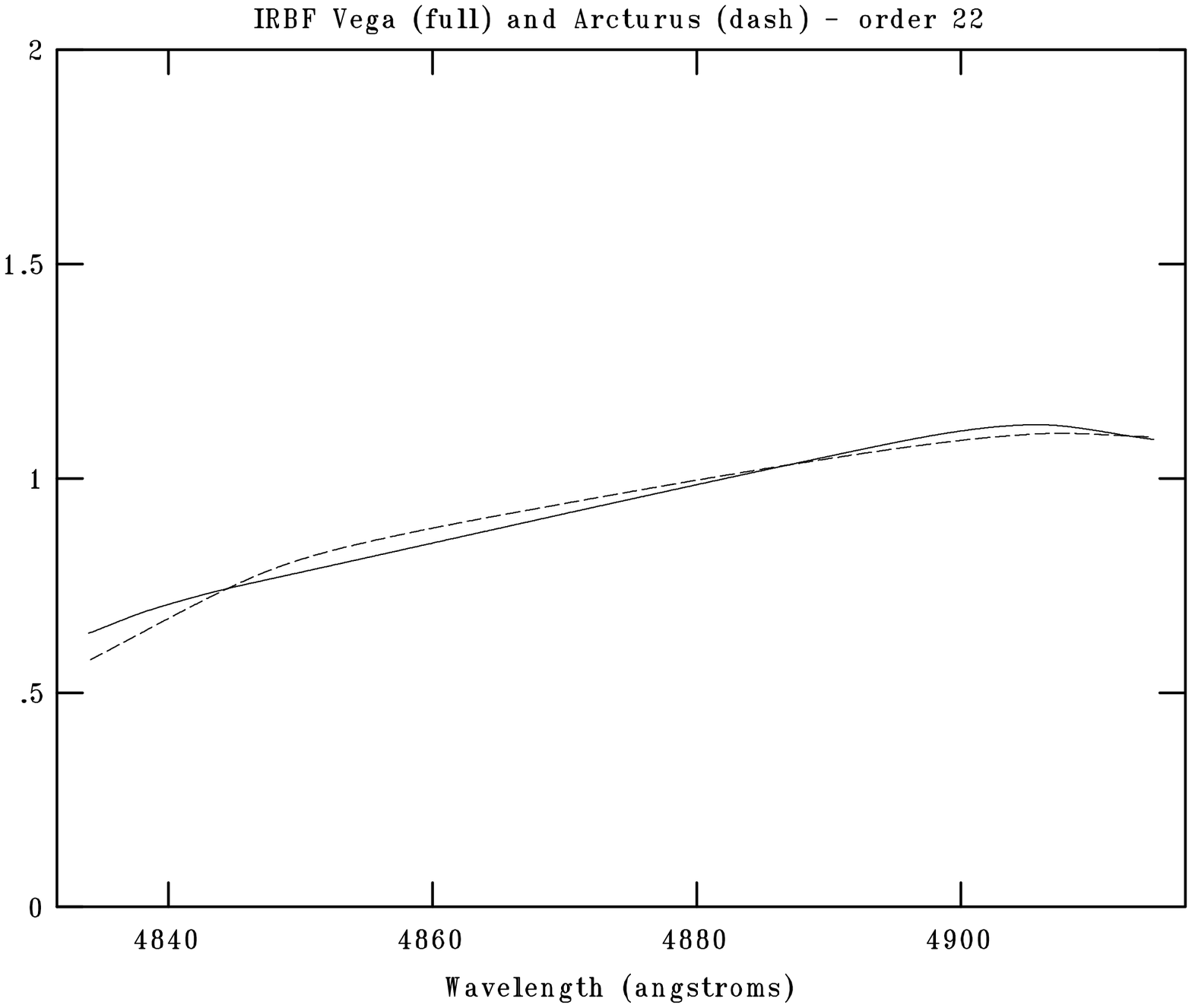}\\
\caption{Comparison of IRBFs in two orders of Arcturus and Vega.
\textbf{Left:} Order 13. 
\textbf{Right:} Order 22.}
\label{IRBFs}
\end{center}
\end{figure}

Our order correction procedure is based on hypothesis supposing the RBF to be
just product of some instrument-dependent part, called Intrinsic Residual Blaze
Function (IRBF) and the GSF, which is observation dependent.  We can get it
from division of RBF by GSF.  Example of IRBF for the two orders shown above
(see Fig.~\ref{RBFord}) is given in Fig.~\ref{IRBFs}.

\begin{figure}[h]
\begin{center}
\includegraphics[width=11.5cm,height=8.5cm]{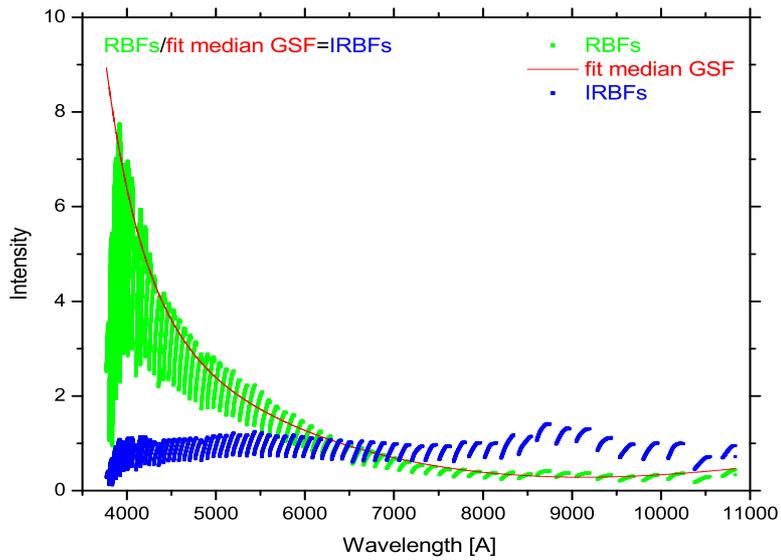}\\
\caption{RBF and IRBF of Vega with median fit of GSF}
\label{vegarbf_GSF}
\end{center}
\end{figure}

\begin{figure}
\begin{center}
\includegraphics[width=11.5cm,height=8.5cm]{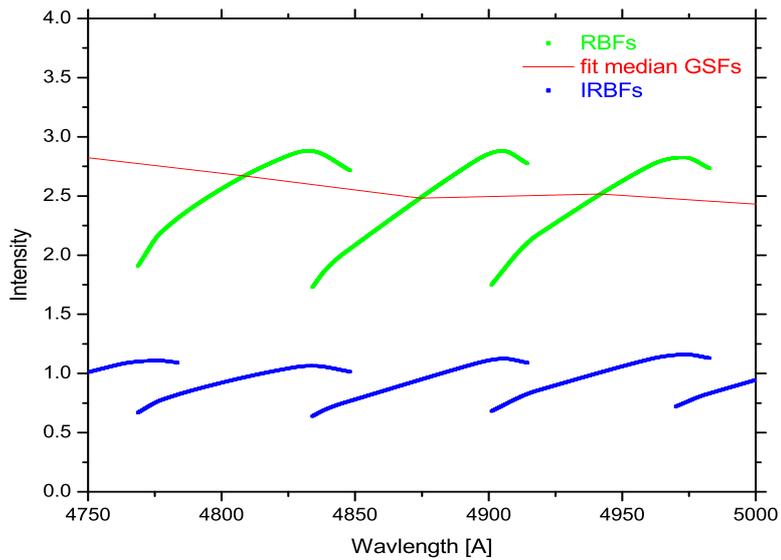}\\
\caption{Detail view on RBF and IRBF on Vega with median fit of GSF}
\label{vegarbf_GSF_detail}
\end{center}
\end{figure}

Once known, the IRBF can be divided out from flat-fielded stellar orders (on
extracted spectra) and the corrected orders should be matching better thus
allowing easy merging of orders.  Some examples of this procedure in global and
detailed view is given on Fig.~\ref{vegarbf_GSF} and
Fig.~\ref{vegarbf_GSF_detail}.  If the GSF for given observation is estimated,
the resulting spectrum should become even continuum normalized (to level 1.0) 

\subsection{Correction of Flat-Fielded Stellar Orders}

\begin{figure}[h]
\begin{center}
\includegraphics[width=11.5cm,height=8.5cm]{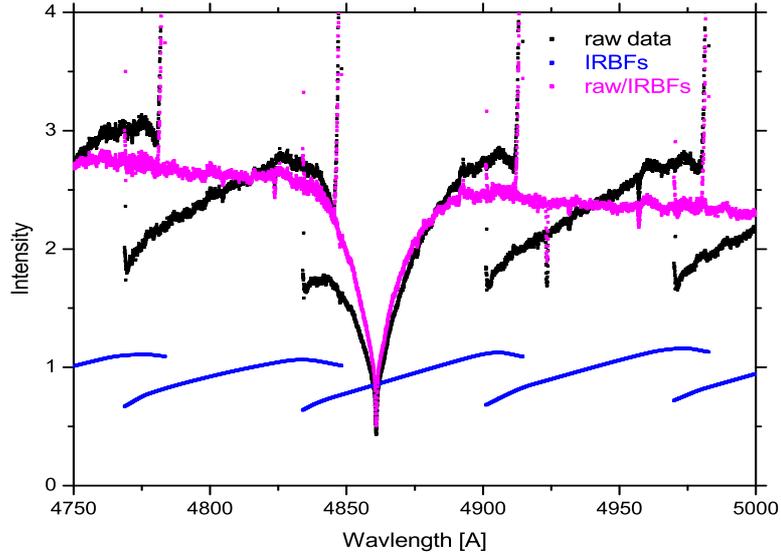}\\
\caption{Raw unblazed data after correction by IRBF. Merged profile of ${H_\beta}$}
\label{corrhbeta}
\end{center}
\end{figure}

The final goal of the reduction process is the continuum normalized (rectified)
merged spectrum of a target.  If the IRBF is really constant for given
spectrograph, we can built it in an reduction pipeline, so every flat-fielded
order is divided by it, before merging in one long spectrum
(Fig.~\ref{corrhbeta}).

\begin{figure}[h]
\begin{center}
\includegraphics[angle=90,width=0.45\textwidth]{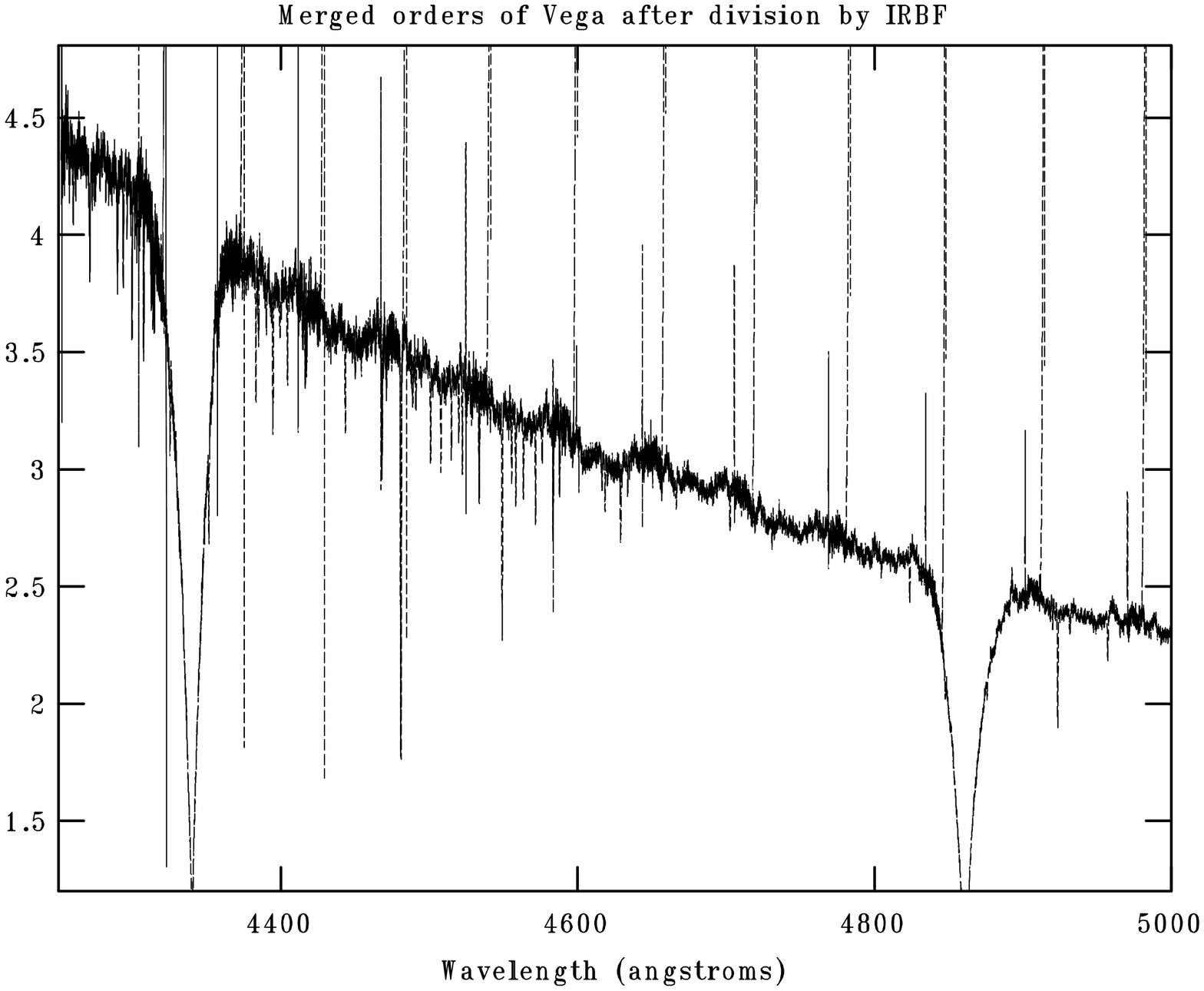}
\includegraphics[angle=90,width=0.45\textwidth]{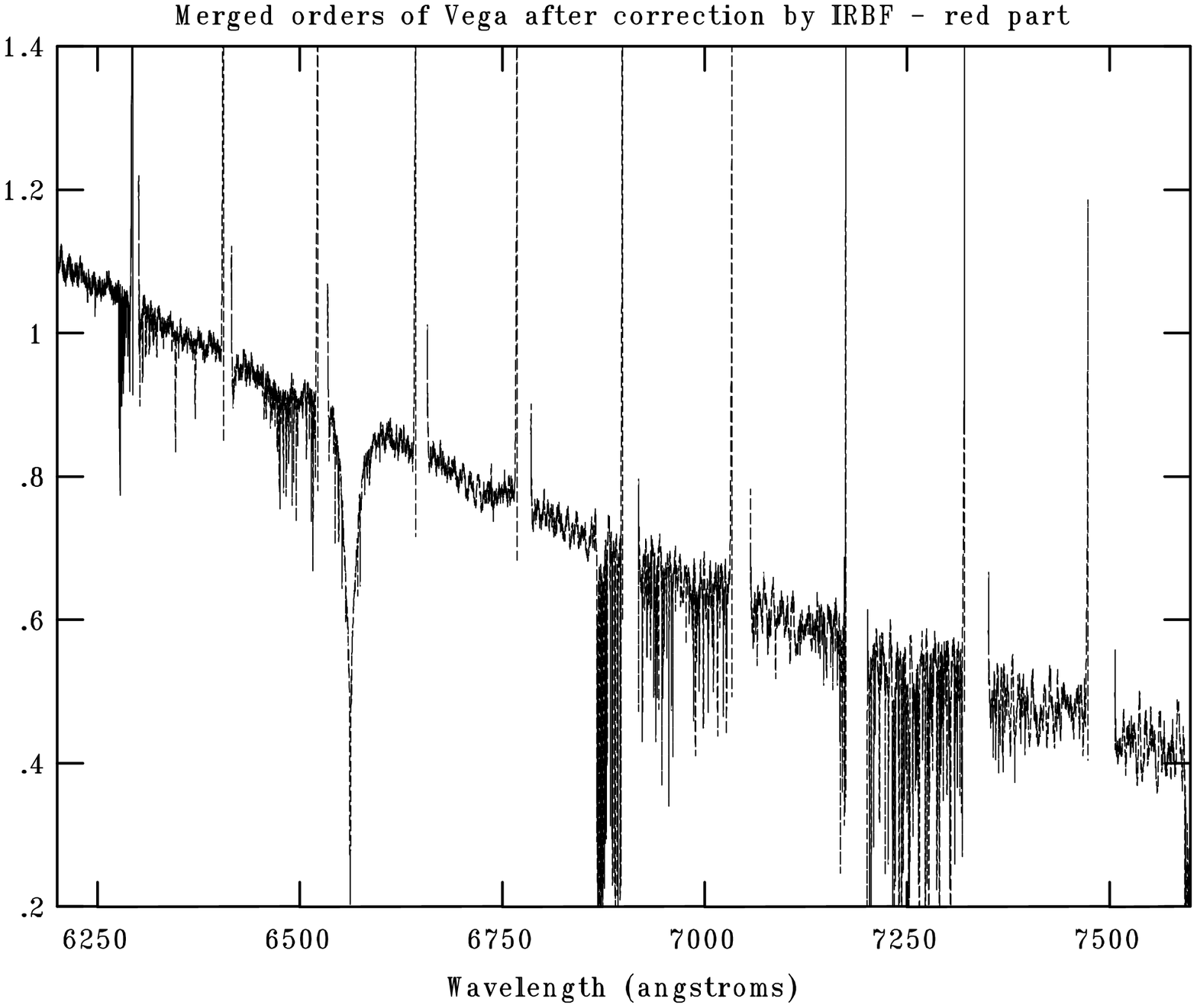}\\
\caption{Merged orders of Vega in different spectral regions. The vertical
lines mark the edges of individual orders. Note the gaps in red region.
\textbf{Left:} Blue region. 
\textbf{Right:} Red region.
}
\label{merged}
\end{center}
\end{figure}

This correction will assure the consistent behavior of order edges in the
region of overlapping orders and thus smooth connection of neighboring orders
together.  See Fig.~\ref{merged} for example of merged Vega orders in different
spectral regions.

\section{Conclusion}

We have shown a one possible way how to tackle the problem of precise unblazing
of echelle spectra using the separation of correction function to
instrument-only dependent part (IRBF) changing on the scale of one echelle
order and observation-dependent part (GSF) smoothly changing over the whole
observed spectrum. 

Even if this is not the final solution, and still some quickly changing
variations remain, the resulting corrected echelle orders may be merged in the
one long spectrum much more easily avoiding strong ripples. If the estimate of
GSF is known for given observation, the spectra may be  normalized to continuum
almost automatically.

Our method is just a first simple approximation of the general correction
procedure, it may be instrument dependent (better suited to slit-fed
spectrographs) and so the future investigation of the problem is still highly
desirable.

\acknowledgments

This research has was supported by grant  205/06/0584 of the Granting Agency of
the Czech Republic. The Astronomical Institute of the Academy of Sciences of
the Czech Republic is   supported by project AV0Z10030501.

We have extensively used the reduction system IRAF  distributed by the National
Optical Astronomy Observatories, which is operated by the Association of
Universities for Research in Astronomy, Inc. (AURA) under cooperative agreement
with the National Science Foundation.  \input mainjournals.tex

\bibliography{apj-jour,echelle,grant2004,grant2005,spie}
\bibliographystyle{spiebib} 

\end{document}

%% file: mainjournals.tex
\def\aj{AJ}%
          % Astronomical Journal
\def\araa{ARA\&A}%
          % Annual Review of Astron and Astrophys
\def\apj{ApJ}%
          % Astrophysical Journal
\def\apjl{ApJ}%
          % Astrophysical Journal, Letters
\def\apjs{ApJS}%
          % Astrophysical Journal, Supplement
\def\ao{Appl.~Opt.}%
          % Applied Optics
\def\apss{Ap\&SS}%
          % Astrophysics and Space Science
\def\aap{A\&A}%
          % Astronomy and Astrophysics
\def\aapr{A\&A~Rev.}%
          % Astronomy and Astrophysics Reviews
\def\aaps{A\&AS}%
          % Astronomy and Astrophysics, Supplement
\def\azh{AZh}%
          % Astronomicheskii Zhurnal
\def\baas{BAAS}%
          % Bulletin of the AAS
\def\jrasc{JRASC}%
          % Journal of the RAS of Canada
\def\memras{MmRAS}%
          % Memoirs of the RAS
\def\mnras{MNRAS}%
          % Monthly Notices of the RAS
\def\pra{Phys.~Rev.~A}%
          % Physical Review A: General Physics
\def\prb{Phys.~Rev.~B}%
          % Physical Review B: Solid State
\def\prc{Phys.~Rev.~C}%
          % Physical Review C
\def\prd{Phys.~Rev.~D}%
          % Physical Review D
\def\pre{Phys.~Rev.~E}%
          % Physical Review E
\def\prl{Phys.~Rev.~Lett.}%
          % Physical Review Letters
\def\pasp{PASP}%
          % Publications of the ASP
\def\pasj{PASJ}%
          % Publications of the ASJ
\def\qjras{QJRAS}%
          % Quarterly Journal of the RAS
\def\skytel{S\&T}%
          % Sky and Telescope
\def\solphys{Sol.~Phys.}%
          % Solar Physics
\def\sovast{Soviet~Ast.}%
          % Soviet Astronomy
\def\ssr{Space~Sci.~Rev.}%
          % Space Science Reviews
\def\zap{ZAp}%
          % Zeitschrift fuer Astrophysik
\def\nat{Nature}%
          % Nature
\def\iaucirc{IAU~Circ.}%
          % IAU Cirulars
\def\aplett{Astrophys.~Lett.}%
          % Astrophysics Letters
\def\apspr{Astrophys.~Space~Phys.~Res.}%
          % Astrophysics Space Physics Research
\def\bain{Bull.~Astron.~Inst.~Netherlands}%
          % Bulletin Astronomical Institute of the Netherlands
\def\fcp{Fund.~Cosmic~Phys.}%
          % Fundamental Cosmic Physics
\def\gca{Geochim.~Cosmochim.~Acta}%
          % Geochimica Cosmochimica Acta
\def\grl{Geophys.~Res.~Lett.}%
          % Geophysics Research Letters
\def\jcp{J.~Chem.~Phys.}%
          % Journal of Chemical Physics
\def\jgr{J.~Geophys.~Res.}%
          % Journal of Geophysics Research
\def\jqsrt{J.~Quant.~Spec.~Radiat.~Transf.}%
          % Journal of Quantitiative Spectroscopy and Radiative Trasfer
\def\memsai{Mem.~Soc.~Astron.~Italiana}%
          % Mem. Societa Astronomica Italiana
\def\nphysa{Nucl.~Phys.~A}%
          % Nuclear Physics A
\def\physrep{Phys.~Rep.}%
          % Physics Reports
\def\physscr{Phys.~Scr}%
          % Physica Scripta
\def\planss{Planet.~Space~Sci.}%
          % Planetary Space Science
\def\procspie{Proc.~SPIE}%
          % Proceedings of the SPIE